\documentclass[aps,prl,twocolumn,groupedaddress,showpacs]{revtex4}
\usepackage{graphicx}
\usepackage[ansinew]{inputenc}
\usepackage{array}
\usepackage{color}
\usepackage{amsmath}
\usepackage{amsxtra}
\usepackage{amssymb}
\usepackage{latexsym}
\usepackage{dsfont}
\begin{document}
\title{Quantum Mechanical Limits to Inertial Mass Sensing by Nanomechanical
Systems}
\author{P.-L. Giscard, M. Bhattacharya and P. Meystre}
\affiliation{B2 Institute, Department of Physics and College of
Optical Sciences, The University of Arizona, Tucson, Arizona
85721}

\date{\today}

\begin{abstract}
We determine the quantum mechanical limits to inertial mass-sensing based on
nanomechanical systems. We first consider a harmonically oscillating cantilever whose vibration frequency is changed by mass accretion at its surface. We show that its zero-point fluctuations limit the mass sensitivity, for attainable parameters, to about an electron mass. In contrast to the case of a classical cantilever, we find the mass sensitivity of the quantum mechanical cantilever is independent of its resonant frequency in a certain parameter regime at low temperatures. We then consider an optomechanical setup in which the cantilever is reflective and forms one end of a laser-driven Fabry-P\'{e}rot cavity. For a resonator finesse of $5$ the mass sensitivity at $T=0$ is limited by cavity noise to about a quarter of a Dalton, but this setup has a more favorable temperature dependency at finite temperature, compared to the free cantilever.
\end{abstract}


\maketitle

Mechanical oscillators are quickly approaching the quantum regime. Advances in
nanofabrication, laser-cooling, electromechanics etc. have allowed experimentalists to cool harmonically oscillating cantilevers to an excitation number of $n \sim 25$ \cite{Schwab2006}, the eventual aim being to reach the ground state $n=0$, see Ref.~\cite{Vahala2008} and references therein. The availability of macrosocopic mechanical systems operating in the quantum regime will allow one to address conceptual questions regarding the nature of the
quantum-classical interface \cite{Penrose2003}, to coherently control atomic
\cite{Hansch2007} and molecular \cite{Swati2008} systems; and to construct displacement \cite{Heidmann2006}, mass \cite{Roukes2004}, and force \cite{Cohadon2007} sensors of unprecedented sensitivity.

A prime example of an ultra-precise displacement sensor is the Laser Interferometer Gravitation-Wave Observatory (LIGO), which is expected to enter the quantum regime in the coming years \cite{Mavalvala2008}. A large body of work, both experimental \cite{Mavalvala2008} and theoretical \cite{Caves1980} has been devoted to investigating the ultimate, quantum mechanical limits to the displacement sensitivity of devices such as LIGO.

Mass sensing is another important application of nanomechanical
resonators useful in physical, chemical and biological contexts \cite{Roukes2004}. Typically the mass to be ``weighed'' $\delta M$ is deposited on a harmonically oscillating cantilever of mass $M \gg \delta M$ , whose resonant frequency $\omega_0$ changes as a result. By successfully detecting that small change $\delta \omega_0$, sensitivities have been reached where a single molecule can be weighed \cite{Yang2006}.

The question that we address in this Letter is to determine the ultimate resolution of nanocantilever-based mass sensing. As is the case for gravitational wave detectors this limit is set by the laws of quantum mechanics. However,
to the best of our knowledge all previous descriptions of cantilever-based mass-sensing have been classical \cite{Roukes2004,Yang2006,Cleland2002,Cleland2005,Ekinci2005}, or have
accounted for noise effects only at high temperature \cite{Yurke2006}. These treatments predict a sensitivity to infinitesimally small masses as the
temperature is lowered $(T \rightarrow 0)$, but of course, this result has to be corrected to account for the effects of quantum noise.  Motivated by the rapid approach of mechanical oscillators to the quantum regime, we present a simple quantum mechanical analysis of mass sensing illustrating the limits set by thermal noise at \textit{low} temperatures. Our results confirm that in the $T \rightarrow 0$ limit, mass sensitivity is limited by purely quantum mechanical effects.

We begin by considering the simple case of a mechanical cantilever of mass $M$ and vibration frequency $\omega_0$ that is driven near resonance by an electrical circuit, typically a phase-locked loop that can detect its frequency shift $\delta \omega_0$ when a mass $\delta M$ is deposited on it (see Ref.~\cite{Roukes2004} for a diagram of the circuit). For mass increments $\delta M \ll M$
we have \cite{Roukes2004}
\begin{equation}
\label{eq:masssensi}
\delta M=\frac{\partial M}{\partial\omega_0}\delta\omega_0,
\end{equation}
where the mass responsivity $\partial M/\partial\omega_0=-2M/\omega_0$
\cite{Roukes2004,Sauerbrey1959}.

\begin{figure}
\includegraphics[width=0.35 \textwidth]{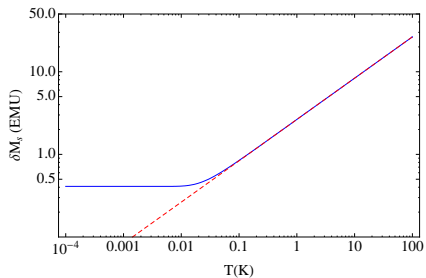}
\caption{\label{fig:mass_sens}(Color online) Quantum (solid curve) and classical
(dashed curve) mass sensitivity $\delta M_{\rm{s}}$ in electron mass units ($1\rm{EMU}\sim9\times10^{-31}$kg)
of the free cantilever mass sensor for $M\sim0.1$fg, $Q=10^{7}$, $\omega_0\sim1$GHz,
$\Delta f\sim1$kHz and $E_{d}\sim1.6\times10^{-15}$J.}
\end{figure}

Various noise sources also couple to the cantilever, resulting in random frequency shifts. In order to obtain a signal to noise ratio (SNR) of at least unity the smallest detectable frequency shift $\delta \omega_{0,\rm s}$ should equal the mean frequency shift due to the various noise sources,
\begin{equation}
\label{eq:orms}
\delta\omega_{0,\rm s}=\left(\int_{0}^{\infty}S_{\omega}(\omega)T(\omega)d\omega\right)^{1/2},
\end{equation}
where $S_{\omega}(\omega)$ is the spectral density of noise-induced frequency
fluctuations and $T(\omega)$ is the normalized transfer function of the circuit realizing the measurement. Following \cite{Roukes2004}, that expression can
be simplified if the measurement circuit responds only to frequencies
in a bandwidth $\Delta f$ around $\omega_0$. Approximating then $T(\omega)$ by an ideal band pass filter with $T(\omega)=1$ between
$\omega_0-\pi\Delta f$ and $\omega_0+\pi\Delta f$ and $0$ elsewhere, Eq.~(\ref{eq:orms}) yields the smallest detectable frequency shift as
\begin{equation}
\label{eq:deltasmall}
\delta\omega_{0,\rm s}=\displaystyle \left(\int_{\omega_0-\pi\Delta f}^{\omega_0+
\pi\Delta f}S_{\omega}(\omega)d\omega\right)^{1/2}.
\end{equation}
The spectral density $S_{\omega}(\omega)$ depends on the specific system noise source being considered. In order to isolate the {\em fundamental}  noise limitations to mass sensing we assume in the following that both the drive and the detection are noiseless, despite the fact that in reality they introduce backaction noise. In practice, transduction needs to be optimized to reduce backaction noise to a level comparable to the intrinsic noise that we consider here \cite{Ekinci2005}. Considering the parameters given in the caption of Fig.~\ref{fig:mass_sens}, we also neglect other technical sources of noise such as temperature fluctuations, which yield limitations to mass sensitivity smaller than one electron mass for temperatures below $1$K \cite{Roukes2004}; elastic and inelastic collisions with background gas particles, whose influence on mass sensitivity also drops below one electron mass for a vacuum of $10^{-6}$ Torr \cite{Roukes2004,Cleland2002}, and defect motion within
the cantilever itself. As pointed out in Ref.~\cite{Ekinci2005}, under these conditions the operational limits of mass sensing are set by \textit{thermomechanical} displacement fluctuations in the cantilever.

With these caveats in mind we treat the cantilever as a quantum harmonic oscillator of mechanical quality factor $Q$ and total thermal noise energy
$E_{\rm noise}=\frac{1}{2}\kappa\langle q^{2}_{\rm noise}\rangle$, where $\kappa =M \omega_0^{2}$ is the stiffness of the cantilever and $\langle q^{2}_{\rm noise}\rangle$ is its center-of-mass mean square displacement. The cantilever is driven on resonance by a noiseless circuit with driving energy $E_{d}$.

From the equation of motion of the cantilever, the spectral density of noise-driven random displacements $S_{q}(\omega)$ is given by
\cite{Heer1972}
\begin{equation}
\label{eq:Sqtot}
S_{q}(\omega)=\frac{2\omega_0}{MQ}\frac{E_{\rm noise}}{(\omega^{2}-
\omega_0^{2})^{2}+\omega^{2}\omega_0^{2}/Q^{2}}.
\end{equation}
This spectral density can be linked to the spectral
density of frequency fluctuations $S_\omega(\omega)$ through the relations
$S_{q}(\omega)=\langle q^2_d\rangle S_\phi(\omega)/(2\pi)^2$ and
$S_\omega(\omega) \simeq \omega_0^2S_\phi(\omega)/(2\pi)^2Q^2,$
where $S_\phi(\omega)$ is the spectral density of phase fluctuations and
$\langle q^{2}_{d}\rangle$ is the mean squared amplitude of the cantilever
resulting from the (noiseless) resonant drive \cite{Robins1982}. Combining these expressions with Eqs.~(\ref{eq:deltasmall}) and ~(\ref{eq:Sqtot}) the smallest mass detectable is given by
\begin{equation}
\label{eq:mass}
\delta M_{\rm s}=\frac{-2M}{\omega_0}\left(\int_{\omega_0-
\pi\Delta f}^{\omega_0+\pi\Delta f}d\omega\frac{\omega_0^{5}}{Q^{3}}\frac{E_{\rm noise}/E_{d}}{\left(\omega^{2}-\omega_0^{2}\right)^{2}+\omega^{2}\omega_0^{2}/Q^{2}}\right)^{1/2},
\end{equation}
where $E_{d}=\frac{1}{2}M\omega_0^{2}\langle q^{2}_{d}\rangle$ is the driving energy.

To evaluate $\delta M_{\rm s}$ explicitly requires an expression for the
noise energy $E_{\rm noise}$. For a quantum harmonic oscillator in
equilibrium with a thermal reservoir at temperature $T$ we have
\begin{equation}
\label{eq:enoise}
E_{\rm noise} =\hbar\omega_0(\bar{n}+\frac{1}{2}),
\end{equation}
where $\bar{n}=\left(e^{\hbar \omega_0/k_{B}T}-1\right)^{-1}$ is the mean number of thermal phonons at frequency $\omega_0$ and temperature $T$. $E_{\rm noise}$ can be easily modified to account for cantilever damping, see e.g. Ref.~\cite{Weiss1984}. However these corrections are negligible for the high quality factors being considered here for mass sensing. At $T = 0$
Eq.~(\ref{eq:enoise}) gives $E_{\rm noise} = \hbar\omega_0/2$, the ground state energy of the cantilever, while at high temperatures $T \gg \hbar \omega_0/k_{B}$, $E_{\rm noise} = k_{B}T$ which yields the classical results
published earlier.

The integral from Eq.(\ref{eq:mass}) has been evaluated analytically in the limit $Q\gg1$ and plotted in Fig.~\ref{fig:mass_sens}. An illuminating form for this integral is obtained with the further approximation $2\pi\Delta f\ll \omega_0/Q$ (this can be achieved using the parameters of Fig.~\ref{fig:mass_sens} but $Q=10^{5}$) simplifying Eq.(\ref{eq:mass}) to
\begin{equation}
\label{eq:quantumdM}
\delta M_{\rm s}=2\sqrt{2\pi}M\left[\frac{\hbar\omega_0(2\bar{n}+1)}{2E_{d}}\right]^{1/2}\left(\frac{\Delta f}{Q\omega_0}\right)^{1/2}.
\end{equation}
As expected, quantum fluctuations prevent the smallest detectable mass $\delta M_{\rm s}$ from going to zero when $T \rightarrow 0$. Rather, it is limited by the quantum zero-point fluctuations of the cantilever to
\begin{equation}
\label{eq:dM}
\delta M_{\rm s}(T=0)=2\sqrt{2\pi}M\left(\frac{\hbar\omega_0}{2E_{d}}
\right)^{1/2}\left(\frac{\Delta f}{Q\omega_0}\right)^{1/2},
\end{equation}
which yields $\delta M_{\rm s}(T=0)\simeq10$ electron masses for our parameters. Note that we have retained $\omega_0$ in Eq.~(\ref{eq:dM}) inside the first parentheses to provide a scale for the driving energy, and inside the second parentheses to facilitate comparison with the corresponding classical expression \cite{Roukes2004}. However these frequency dependencies cancel each other in the quantum regime, indicating that in contrast to the classical regime mass sensitivity is now independent of the mechanical frequency of the cantilever within the range of validity of our description.

Examples of quantum and classical mass sensitivities are plotted as a function of temperature in Fig.~\ref{fig:mass_sens}. For the parameters of that figure, we have $\delta M_{\rm{s}}(T=0)\sim 5\times10^{-31}\rm{kg}$, or about an electron mass.
This limit is reached when $\bar{n}\ll1$, which occurs for
$\omega_0\sim1$GHz at $T<10$mK. Such temperatures have recently been reached experimentally for mechanical oscillators using laser cooling techniques \cite{Thompson2008}. These same techniques also hold the promise of mechanical squeezing of the cantilever motion that would result in
$E_{\rm noise}< \hbar\omega_0/2$ \cite{BhattacharyaSqueeze2007,Zoller2009}
with an associated improvement in $\delta M_{\rm s}$.

We now turn to an analysis of the mass sensitivity afforded by a typical
optomechanical cooling configuration where the vibrating cantilever forms one end of a Fabry-P\'{e}rot cavity driven by laser radiation. The cantilever itself is driven as before by an external noiseless circuit with energy $E_{d}$. Let $p$ and $q$ be respectively the momentum and displacement operators of the cantilever with $[q,p]=i\hbar$. In terms of the dimensionless variables $\tilde{p}=p/\sqrt{\hbar M\omega_{o}}$ and $\tilde{q}=q\sqrt{M\omega_{o}/\hbar}$, the Hamiltonian that models that cavity mass sensor is
\begin{equation}
\label{eq:Ham2}
    H = \hbar \omega_l a^{\dagger}a+\frac{\hbar\omega_o}{2}
    \left(\tilde{p}^{2}+\tilde{q}^{2}\right)-\hbar g a^{\dagger}a\tilde{q}-E_{d}(t)\tilde{q}.
\end{equation}
where $a$ and $a^\dagger$ are bosonic annihilation and
creation operators for the cavity mode of frequency $\omega_l$, $\omega_0$ is the natural vibration frequency of the cantilever and
$g=(\omega_l/L)(\hbar/M\omega_0)^{1/2}$ is the optomechanical
coupling parameter \cite{VitaliVib2007}.

To find the displacement fluctuations of the micro-mirror due to noise, we
evaluate semiclassically $\langle q^{2}_{\rm noise}\rangle$ \cite{VitaliVib2007}, and insert it in the expression $E_{\rm noise}=\frac{1}{2}\kappa_{\rm eff}\langle q^{2}_{\rm noise}\rangle$, with $\kappa_{\rm eff}$ the effective stiffness of the micro-mirror modified by radiation pressure. Specifically, we first calculate the evolution of the operators describing the dynamics of the intracavity field and of the cantilever via a set of quantum Langevin equations (QLE) that include the hamiltonian dynamics of Eq.~(\ref{eq:Ham2}), input laser noise, cavity damping, and the Brownian noise associated with the coupling of the cantilever to a reservoir at temperature $T$
\begin{eqnarray}
\label{eq:QLErot}
\dot{a}&= &-i(\delta-g\tilde{q})a-\frac{\gamma}{2}a+\sqrt{\gamma}a^{\rm in},\nonumber \\
\dot{\tilde{q}}&=& \omega_0\tilde{p},\nonumber \\
\dot{\tilde{p}}&=& -\omega_0\tilde{q}+g a^{\dagger}a+
\frac{E_{d}(t)}{\hbar}-\gamma_{c}\tilde{p}+\xi^{\rm in}.
\end{eqnarray}
Here $\delta =\omega_c-\omega_l$ is the detuning of the laser frequency $\omega_l$ from the cavity resonance $\omega_{c}$, $\gamma$ is the cavity damping rate and $\gamma_{c}$ is the damping rate of the cantilever. The noise operator $a^{\rm in}$ describes the laser field incident on the cavity. Its mean amplitude is $\langle a^{\rm in}(t)\rangle =\alpha_{\rm s}^{\rm in}$ and its
fluctuations are taken to be delta-correlated,
$\langle \delta a^{\rm in}(t) \delta a^{\rm in,\dagger}(t') \rangle=\delta (t-t')$. The
Brownian noise operator $\xi^{\rm in}$ describes the thermomechanical noise that couples to the cantilever from the environment. Its mean value is zero and, for high mechanical quality cantilevers such as considered here, $\omega_0/\gamma_{c} \gg 1$, its fluctuations are delta-correlated $\langle \delta \xi^{\rm in}(t) \delta \xi^{\rm in}(t') \rangle=\gamma_{c}(2\bar{n}+1)\delta(t-t').$

\begin{figure}
\includegraphics[width=0.35\textwidth]{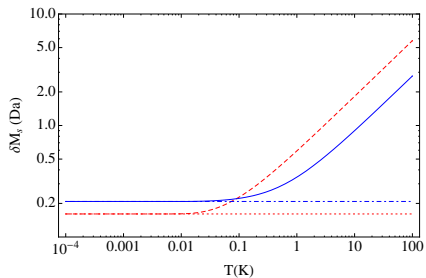}
\caption{\label{fig:mass_sens2}(Color online) Cavity mass sensor mass sensitivities in Dalton units (1Da$\sim1.7\times10^{-27}$kg) as a function of
temperature for the parameters of the text and $P=100\mu$W (dashed red line) and $P=1$mW (solid blue line)
and $\mathcal{F}=5$. The dotted and dot-dashed lines are
the mass sensitivities at $T=0K$ respectively for $P=100\mu$W and $P=1$mW.}
\end{figure}

The semiclassical, steady-state solutions of Eqs.(\ref{eq:QLErot}) oscillate due to the time-dependent driving of the mirror. However the amplitude of these oscillations is small as compared to their mean value (for example with the parameters given in the caption of Fig.~\ref{fig:mass_sens2}, the mean power inside the cavity is about 1.1mW with an oscillation amplitude of the order of $100$nW) about that value. We therefore assimilate the steady state solutions to their temporal mean values which are readily obtained as
$\tilde{p}_{\rm s}=0$ and $\tilde{q}_{\rm s}=g|\alpha_{\rm s}|^2/\omega_0$. The
mean steady state field amplitude $\alpha_{\rm s}$ is complex in general, but it is possible to choose its phase such that $\alpha_{\rm s}$ is real, i.e. $\alpha_{\rm s}=\sqrt{\gamma}|\alpha_{\rm s}^{\rm in}|/\left[(\frac{\gamma}{2})^{2}
+(\delta-g\tilde{q}_{\rm s})^{2}\right]^{1/2}$.
 This amplitude can be bistable for high incident fields \cite{dorsel1983} and appropriate cavity-laser detunings, but one can easily choose system parameters that avoid this regime.

To account for the quantum fluctuations $\delta a$ about the semiclassical
steady-state $\alpha_{\rm s}$ we decompose $a=\alpha_{\rm s}+\delta a$, and similarly for the other operators of Eq.~(\ref{eq:Ham2}). Linearizing the quantum Langevin equations of
motion in the fluctuations gives then
\begin{equation}
\dot{v}(t)=A v(t)+v(t),
\end{equation}
where $v(t)$ is the vector of fluctuations.
Recasting the optical fields in terms of their quadratures $X_a$ and $Y_b$ with fluctuations  $\delta X_{a} =(\delta a+\delta a^{\dagger})/\sqrt{2}$, $\delta Y_{a} =(\delta a-\delta a^{\dagger})/i\sqrt{2}$], and similarly for the input field, the input noise vector becomes $n(t)=(0,\delta \xi^{\rm in},\sqrt{\gamma}\delta X_{a}^{\rm in},\sqrt{\gamma}\delta
Y_{a}^{\rm in})$ and
\begin{equation}
\label{eq:RHmatrix} A =
\begin{pmatrix}
0 &  \omega_0             &  0                    &       0   \\
 - \omega_0            & -\gamma_{c}  & G   &       0   \\
0               &    0             &     -\gamma/2                     &  \Delta\\
G &    0     & -\Delta &-\gamma/2\
\end{pmatrix},
\end{equation}
where $\Delta=\delta-g\tilde{q}_{\rm s}$ is the effective cavity detuning and
$G=g_{\rm s}\sqrt{2}$ is the effective optomechanical parameter.
We have verified that the Routh-Hurwitz criterion is satisfied here, that is, the eigenvalues of $A$ all have a negative real part which ensures the stability of the steady-state solution.

We can determine $\langle q^{2}_{\rm noise}\rangle =\hbar \langle \tilde{q}^{2}\rangle/{M\omega_0}$ from the correlation matrix $C$, whose elements are given by $C_{ij}=[\langle v_i(\infty)v_j(\infty)+v_{j}(\infty)v_{i}(\infty)\rangle]/2.$
This matrix can be shown to be the solution of the equation
$AC+CA^{T}=-D$, with $D=\mathrm{diag}[0,\gamma_{c}(2\bar{n}+1),\gamma/2,\gamma/2]$ \cite{VitaliVib2007}.
The full solution of this equation can be obtained analytically, but it is cumbersome and will not be reproduced here. For our present purpose, we only need to consider the first diagonal element of the matrix $C$ as it yields $\langle \tilde{q}^{2}\rangle$. From this, the integral of Eq.(\ref{eq:mass}) has been evaluated analytically in the limit $Q\gg1$. It is illustrated in Fig.~\ref{fig:mass_sens2} for the same parameters as in Fig.~\ref{fig:mass_sens} except $M=50$fg (1$\mu$m$\times200$nm$\times150$nm silicon cantilever), a cavity length of $L\sim$1mm, $\lambda_{\rm Laser}=810$nm, a cavity resonance frequency $\omega_c\sim2\pi \times 10^{14}$Hz and effective cavity detuning $\Delta=\omega_0$. The laser power $P$ and the finesse $\mathcal{F}$ are given in the caption. Note that with the further approximation $2\pi\Delta f\ll \omega_{0}/Q$ and with Eq.(\ref{eq:mass}) we obtain a simplified expression for the mass sensitivity
\begin{equation}
\delta M_{\rm s}=2\sqrt{\pi}M\left(\frac{\hbar\omega_0\langle \tilde{q}^{2}\rangle}{E_{d}}
\right)^{1/2}\left(\frac{\Delta f}{Q\omega_0}\right)^{1/2}.
\end{equation}

For a cavity finesse $\mathcal{F}=5$, the level of noise introduced by the cavity setup equals the noise due to the vacuum fluctuations of the cantilever. The mass sensitivity is therefore reduced by a factor of two from that of the free cantilever mass sensor when it is a part of this optical arrangement.
One attractive feature of the cavity mass sensor is the weak dependence of its sensitivity on temperature. Indeed, although increasing the input laser power increases the level of noise at $T=0$K as more light is present in the cavity, it decreases the temperature dependence of the mass sensitivity, so that for $P=1$mW the mass sensitivity at $T=25K$ is only about $600\%$ worse than it is at $T=0K$. This is to be contrasted to the situation for the free cantilever mass sensor for which the mass sensitivity degrades by a factor of 30 over the same temperature range. This drastically different behavior is a result of the active cooling by the laser which maintains the micro-mirror at an effective temperature much smaller than that of the bath \cite{BhattacharyaCoolTrap2007}.

On the other hand, one drawback of the cavity mass sensor is the small mass required of the micro-mirror. Eq.(\ref{eq:mass}) shows that the smallest  detectable mass change is proportional to the mass of the cantilever. For our parameters it follows that a cavity sensor mass of $M\sim0.1\rm{fg}$ is needed for $\delta M_{\rm s}$ to approach an electron mass. We also find that the cavity mass sensor requires a cavity finesse approaching $\mathcal{F} \simeq 5$ to reach a regime in which the mass sensitivity is limited by vacuum fluctuations of the cantilever,  implying a reflectivity $R \gtrsim 30\%$. It might be difficult to combine both characteristics -- small mass and high reflectivity -- in one cantilever and we have therefore considered a $50$fg, 1$\mu$m$\times200$nm$\times150$nm silicon cantilever for the cavity based mass sensor. Future work we will study whether a three-mirror configuration presents additional advantages \cite{ThompsonExp3MC2008, BhattacharyaCoolTrap2007}. 

As a final remark, we note that more advanced detectors based on squeezed cantilever motion will only be attractive if the level of noise introduced by the optical setup that produces such nonclassical cantilever states \cite{BhattacharyaSqueeze2007,Zoller2009} drops below the noise level due to vacuum fluctuations of a cantilever in a coherent state (that is for $\mathcal{F}>5$). An alternative might  be to turn off the laser when the cantilever motion is squeezed. This would require stroboscopic mass
measurements which would in turn compete with the requirements on the value of $\Delta f$ needed to obtain good mass sensitivities (here $\Delta f\sim$1kHz).

In summary, we have determined the quantum limit to nanocantilever-based mass sensing to be of the order of an electron mass for realizable parameters; we also note that the mass sensitivity of the quantum cantilever, as opposed to its classical counterpart, can be independent of its frequency at low temperatures. Our findings are of use to high sensitivity detection of isotopes and stoichiometry of compounds. It will also be intriguing to explore the consequences of our findings for spin, charge and force sensing.

\begin{acknowledgments}
This work is supported in part by the US Office of Naval Research, by the National Science Foundation and by the US Army Research Office.
\end{acknowledgments}
\bibliography{Mass_sensor}
\end{document}